\documentclass[twocolumn]{article}

\usepackage{cite}
\usepackage{amsmath}
\usepackage{amsfonts}
\usepackage{algorithmic}
\usepackage{array}
\usepackage{hyperref}
\usepackage{cancel}

\usepackage{fixltx2e}
\usepackage{graphicx}
\usepackage{url}
\usepackage{xcolor}
\usepackage{float}

\newcommand{\myhref}[2]{\textcolor{blue}{\href{#1}{#2}}}
\newcommand{\myurl}[1]{\textcolor{blue}{\url{#1}}}

\newcommand{\ignore}[1]{}

\begin{document}
\title{Circular Programs and Self-Referential
  Structures\footnote{
    first version November 1987; a later version
    Software Practice and Experience, 19(2), 99-109, 1989.
  }
}
\author{ Lloyd Allison, \\
  Department of Computer
  Science\footnote{later the Faculty of Information Technology (FIT)}, \\
  Monash University, Clayton, Victoria 3800, Australia \\
  lloyd.allison@monash.edu
}

\date{}

\maketitle

\begin{abstract}
A circular program creates a data structure whose computation depends
upon itself or refers to itself.
The technique is used to implement the classic data structures
circular and doubly-linked lists, threaded trees and queues,
in a functional programming language.
These structures are normally thought to require updatable variables
found in imperative languages.
For example, a functional program to perform
the breadth-first traversal of a tree is given.
Some of the examples result in circular data structures when evaluated.
Some examples are particularly space-efficient by avoiding the creation
of intermediate temporary structures which would otherwise later become garbage.
Lastly, the technique can be applied in an imperative language
to give an elegant program.

\end{abstract}

\section{Introduction}
\label{sec:Intro}

Bird \cite{Bir84} describes the use of \textit{circular programs}
and applies the technique to transform a program making multiple passes over a
data structure into another program making only one pass.
He attributes knowledge of the technique to Hughes and to Wadler.
Bird's examples are specialized and given in a program-transformation setting.
The purpose of this paper is firstly to show that circular programs
are more widely applicable as a programming technique.
Under suitable circumstances,
circular programs can be used to program
circular and doubly linked lists,
threaded trees and queues \cite{All87}.
These classic data structures are normally thought to require updatable
variables.
Secondly, a circular program can be more space-efficient
than a conventional program by avoiding the creation
of temporary data structure which need to be garbage collected later.
Examples include removing duplicates from a list
and variations on the sieve of Eratosthenese.
Lastly, a circular program can often be translated into
an imperative language, if that is necessary,
giving an elegant and efficient program.

A circular program involves a recursive or self-referential expression
for a data structure.
\begin{verbatim}
  let rec ds = f(ds)
  in      ds
\end{verbatim}

Note that `\texttt{ds}' is a data structure and not a function.
To write a circular program in a functional language requires
a lazy language \cite{FW76,Hen76}.
The evaluation of the data-structure refers to
the data-structure itself; this plainly rules out a strict language.
The evaluation must not use any part or attribute of the structure
before it has been, or can be, computed as this would call for prescience.
Many applications involve specifying a structure before its contents are known
and this is a \textit{forte} of lazy languages.

There appears to be no universally accepted
and precise definition of lazy evaluation.
The bare minimum for the programs in this paper to run correctly
is a call by name mechanism
for binding the right-hand sides of local definitions and
for passing parameters.
This would however be enormously inefficient and result in duplicate
evaluation of many expressions and void the point of the exercise.
The minimum requirement to gain the full advantages of circular programs
is a graph reduction mechanism
using call by need for
the right hand side of local definitions, for parameter passing
and in particular for the parameters of type constructors
such as \textit{cons} or `.' for lists.
Under call by need, an object is bound to a recipe \cite{Hen80} which will
produce the desired value if and when it is needed.
If the value is needed then the recipe is forced or evaluated and
the recipe is also overwritten so that the value is immediately
available thereafter.
A value may contain sub-recipes and these are not forced until necessary.
If a value contains a recursive reference to itself,
this is implemented as a pointer to the top of the value -
hence graph reduction.
All of this is invisible to the programmer.
Lazy evaluation is taken to mean such a system throughout this paper.

In the scheme above, \texttt{ds} is bound initially to a
recipe for \texttt{f(ds)}.
As parts of \texttt{ds} are used in other computations the recipe
is progressively evaluated.
When \texttt{f} finally makes reference to
\texttt{ds} all is well if the required
parts and attributes have been computed.
If \texttt{f(ds)} simply incorporates \texttt{ds} itself in its value,
\texttt{ds} must already have been forced to avoid non-termination.
In this case a pointer to the ``top'' of the data structure
is incorporated and a circular
data structure results in the underlying implementation
of the language.
There is then a path from the top of the data structure
only through type constructors,
or unevaluated functors in logic-programming terms,
back to the top of the structure.
This can be very efficient as some infinite values can be
represented in finite space!
In other examples \texttt{f} uses \texttt{ds} in some other way and no
circular data structure is created.

The programs given here do not require the even more parsimonious
evaluation rule called full laziness (although it does no harm)
which guarantees that no expression at all is evaluated twice.
As an example \cite{Pey85} consider:
\vspace{3\textheight}  
\begin{verbatim}
  let    f x y = sqrt x + y
  in let g = f 4
  in     g 1 + g 2
\end{verbatim}
the expression \texttt{sqrt\ x = sqrt\ 4 =\ 2} is evaluated twice.
In a fully lazy system
an optimization equivalent to the following is made:
\begin{verbatim}
  let    f x = let sqrtx = sqrt x
               in lambda y. sqrtx + y
  in let g = f 4    -- = lambda y.2+y
  in     g 1 + g 2
\end{verbatim}
and $sqrt{x}$ is only evaluated once.

A final requirement for writing any functional program
is that the data structure, \texttt{ds}, be of the single assignment type.
That is, values are not changed once they are known.
Many uses of data structures do have this property.
Sometimes the fact is disguised in an imperative coding
in that a value may be tagged,
set to \texttt{nil} or otherwise marked until the proper value is known.

In the following sections various examples are
given of circular programs in a functional language.
The use of the technique in imperative languages is then discussed.

\section{Functional Examples}

Various applications of circular programs follow.
A simple functional language is used
in which local definitions are included by
`\texttt{let\ in}' or by `\texttt{where}'.
Recursive definitions are qualified by `\texttt{rec}'.
Lists are frequently used and the empty list
is denoted by \texttt{nil} or by `\texttt{[]}' and
the list constructor (\texttt{cons}) by\ `\texttt{.}'.
%

\subsection{Circular Lists}

The simplest circular program of all creates the apparently
infinite list \texttt{[1, 1, ...]}.
\begin{verbatim}
  let rec ones = 1.ones
  in      ones
\end{verbatim}

This is easily evaluated in a lazy language and the implementation
creates a circular list containing one cell that points to itself.
Initially \texttt{ones} is bound to a recipe for 1.\texttt{ones}.
Assuming that this is evaluated, a list cell is created
which contains recipes for 1 and for \texttt{ones}.
If the latter is evaluated, it is also overwritten --
with the value of \texttt{ones} which is a pointer to the list cell.
This creates the following graph:
\begin{verbatim}
  ones:---------> . <-----
                 / |     ^
                /  |     |
               /   --->--|
              1
\end{verbatim}

A program scheme for creating many, although by no means all, circular lists
generalizes \texttt{ones}:
\begin{verbatim}
  let circ x = c
    where rec
      c = build x  -- c self-referential
    and build y =
      f(y).(if p y then c else build (g y))
\end{verbatim}

\texttt{p} is some predicate and \texttt{f} and \texttt{g}
are arbitrary functions.
Note that \texttt{c} is a data structure and \texttt{build} is a function.
The result is a list, \texttt{c} which is equal to
the result of appending
\texttt{[f(x), f(g(x)), f(g}\textsuperscript{2}\texttt{(x))
 ..., f(g}\textsuperscript{n}\texttt{(x))]}
and \texttt{c}
where $n \ge 0$ and
\texttt{p(g}\textsuperscript{n}\texttt{(x))} is true.
The final \texttt{c} is implemented as a
pointer back to the start of the list.
It is possible to eliminate \texttt{c} by
substituting it in build to get the following program:
\begin{verbatim}
  let uncirc x = build x
    where rec
      build y =
        f(y).(if p y then build x
                     else build (g y))
\end{verbatim}

This produces the correct value but it is no longer implemented
as a circular structure;
the data structure is unfolded:
\texttt{[f(x), ..., f(g}\textsuperscript{n}\texttt{(x)),
f(x), ..., f(g}\textsuperscript{n}\texttt{(x)),
f(x),...]}.
This is an equivalent, although much more wasteful, way of representing the
value.
Note that a system using string reduction rather than graph reduction
is liable to produce this data structure for program circ.
Hughes \cite{Hug85} describes a mechanism called
\textit{lazy memo functions} which
would build the circular data structure for the program uncirc
by remembering and reusing function results for past inputs --
such as \texttt{build\ x}.

\subsection{Doubly Linked Lists}

A doubly linked list can be defined in a manner similar to a circular list.
A doubly linked list is either \texttt{nil} or contains three things --
a pointer to a predecessor,
an element and a pointer to a successor node.
\begin{verbatim}
 datatype dbl_list = nil |
     dbl of dbl_list * elt_type * dbl_list

 let double x = build [] x
   where rec
     build prev y =
       if p y then []
       else d
         where rec
           d = dbl(prev, f y, build d (g y))
\end{verbatim}

If \texttt{p\ y}  is immediately true
build returns \texttt{nil}
otherwise it creates a node \texttt{d}.
The node points to its predecessor \texttt{prev}.
It also points to the successor created by a recursive call of \texttt{build}.
The predecessor of this next node is \texttt{d}.
texttt{d}\ is local to build because the predecessor of a node was created
by the preceding call to \texttt{build}.
On the other hand, \texttt{c} is global to
the routine for circular lists because
the start point remains the same through the recursive calls.

In an imperative language a doubly-linked list would be created
one node at a time.
The successor pointer of a node would be set to \texttt{nil}
or left undefined until the successor was created.
The pointer would then be overwritten.
In the circular program above,
a node is created although part of it (the successor pointer) is unevaluated.
The node can still be passed as a parameter so that a
pointer to it can be included as the predecessor of the succeeding node.

Given a lazy language, any amount of scanning the doubly linked list
backwards and forwards causes no extra copies of the list
to be created.
\texttt{d}\ is directly recursive and can only be removed
first by using a fixed-point operator and then
by substituting
in \texttt{build} but then, as in the previous example,
no circular data structure is created.

\subsection{Threaded Trees}

A node of a binary tree contains an element
and pointers to the left and right subtrees.
Most of the tree consists of leaves and most of the pointers are empty.
A \textit{threaded tree} \cite{PW73}
uses those right pointers that would be empty
to point to successor nodes in infix order.
The threads allow the elements in the tree
to be processed sequentially by an iterative or linear-recursive routine.
\begin{verbatim}
  datatype threaded_tree = empty |
    thrd of threaded_tree |
    fork of threaded_tree *
            elt_type *
            threaded_tree
\end{verbatim}
Such a tree may be created in an imperative language
by overwriting empty pointers when the threads were known.

Provided that all the elements to be placed in the tree are given at one time,
a circular program can be written to create a threaded tree.
(In this example the tree is also a binary search tree without duplicates.)
\begin{verbatim}
  let thread L = build true empty L
    where rec
      build isleft succ L =
        if null L then
          if isleft then empty
          else thrd(succ)
        else t
          where rec t =
            fork(build true t
                   (filter (< hd L) L),
                 hd L,
                 build false succ
                   (filter (> hd L) L))
\end{verbatim}
The input elements are in the list \texttt{L}.
\texttt{filter} is a common function to
select elements from a list according to
a predicate or test:
\begin{verbatim}
  let rec filter p l =
     if null l then []
     else let h=hd l
          and rest = filter(tl l)
          in  if p h then h.rest else rest
\end{verbatim}
If \texttt{L} is not \texttt{null}, a node \texttt{t} is built.
\texttt{t}\ contains a left subtree and
either a right subtree or a thread.
The successor thread for the left subtree is \texttt{t} itself.
The successor thread for the right subtree is the successor of \texttt{t}.
This example uses only right threads but left or
predecessor threads are easily added.
The requirement that all elements be given in a list\ \texttt{L}
is to ensure that the tree does not need to be updated
when new elements arrive.

\subsection{Breadth-First Traversal}

The next example and following ones
use expressions which are recursive or circular
but whose values, the result of evaluation, are not
and thus no circular data structures are created.

Prefix, infix and postfix traversals of a tree are easily
programmed in a functional language
but breadth-first traversal is harder.
The usual imperative algorithm employs a queue
and seems to need destructive assignment.
An element is taken from the front of the queue for traversal and
its children are added to the end of the queue.
This appears to imply that either
the queue must be updated or that new copies of a modified
queue must be created
at each step.
A circular program can be written however
in which elements are removed from the front of the queue
as the end is still being computed:
\begin{verbatim}
datatype tree = empty
              | fork of tree * elt_type
                             * tree

let bfirst t = r  -- bfirst: tree->list
 where rec
  r = case t of
      empty => [] |
      fork(left,elt,right) => t.(bf r 1)
 where rec
  bf q n =         -- bf: list->int->list
   if n=0 then []  -- q is used up
   else let root = hd q
    and rest = bf (tl q)
    in case root of
       fork(empty,e,empty)=>rest(n-1)  |
       fork(left, e,empty)=>left.(rest n) |
       fork(empty,e,right)=>right.(rest n) |
       fork(left, e,right)=>
             left.(right.(rest(n+1)))
\end{verbatim}

\texttt{bfirst} returns a list or queue
of the nodes of the tree \texttt{t} in breadth-first order.
If \texttt{t} is not empty, the first node is the root \texttt{t} itself;
at this point the queue has one known element.
\texttt{bf}\ is the central function.
It absorbs a queue \texttt{q} whose known part is
of length \texttt{n} while computing the
result queue; the two queues are in fact one.
\texttt{n}\ indicates the shape of that part of the structure \texttt{r}
that can be used safely.
In an imperative language \texttt{n} would not be necessary
and \texttt{r} might be temporarily terminated by \texttt{nil}
but that cannot be done in a functional language.
\texttt{bf}\ places non-empty children
in the result queue and adjusts its length accordingly;
each call to \texttt{bf} uses one element from \texttt{q} and
adds 0, 1 or 2 elements.
\texttt{rest}\ is a function to build the result literally for the rest of the
input queue after the current node.
Note that \texttt{bfirst} can traverse even infinite trees.
It uses one list cell for each node that is traversed.

\subsection{Unique}

The previous examples illustrated the use of circular programs
to implement classical data structures in functional programs.
The next example uses the technique
to make a space-efficient program.

Consider the problem of writing a function \texttt{unique}.
It is to accept a list as parameter and to return
a list with the same members but with duplicate members
removed and the order of first occurrence is to be maintained.
The problem is close to finding the union of two sets
represented by lists.

It is easy to write such a function if the constraint on order
is dropped:
\begin{verbatim}
  let rec uniqueL L =
    if null L then []
    else if member (hd L) (tl L) then
      uniqueL (tl L)
    else (hd L).(uniqueL (tl L))
\end{verbatim}

\texttt{uniqueL} preserves the order of \textit{last} occurrence so
\texttt{reverse $\circ$ uniqueL $\circ$ reverse}
would solve the original problem.
It would also create garbage in the shape of two intermediate lists.

Another solution to the problem that
is not quite good enough is
\begin{verbatim}
 let rec uniqueF L =
  if null L then []
  else (hd L).(uniqueF (filter (!= hd L) L))
\end{verbatim}
\texttt{uniqueF} certainly preserves the order of first occurrence
but each call of \texttt{filter} creates a temporary list.
\texttt{uniqueF}\ creates $O(|L|)$ such lists and uses $O(|L|^2)$ space.

An imperative programmer might arrive at the following informal description
of a solution.
Unique should create a list \texttt{r}.
It should examine the input \texttt{L}, element by element.
If the current element of \texttt{L} is in \texttt{r}
it should not be added again.
If it is not in \texttt{r} it should be added to \texttt{r}.
There is no need to use an imperative language to implement
this algorithm.
A circular program can use the list \texttt{r}
while creating it at the same time:
\begin{verbatim}
  let unique L = r
    where rec
       r = u L 0
    and u L n =
         if null L then []
         else if member (hd L) r n then
           u (tl L) n
         else (hd L).(u (tl L) (n+1))
    and member e L n =
          if n=0 then false
          else if e=hd L then true
          else member e (tl L) (n-1)
\end{verbatim}
\texttt{r}\ is the self-referential data structure
that function \texttt{u} both creates and uses at the same time.
\texttt{member} is a variation on the conventional list membership function.
While the result \texttt{r} is being built its end is unknown; it
terminates in a recipe.
\texttt{member} cannot therefore use \texttt{null(L)}
to detect the current end point of the
search list.
As in breadth-first traversal,
an integer parameter \texttt{n} is added to keep track of
the length of the known
part of \texttt{r};
it stops \texttt{member} from forcing the recipe and causing an infinite loop.
Note that the shape of a list is represented by a single integer
but that the shape of a tree is more expensive to represent;
a circular program that computes a tree where the computation
depends on the shape of the part already evaluated is
unlikely to be so efficient.

Note that \texttt{uniqueF} and \texttt{unique} will operate on infinite lists
but that only \texttt{unique} creates no intermediate lists and runs
in space linear in the amount of output.

\subsection{Primes}

The final functional examples of circular programs are
variations on the sieve of Eratosthenese.
A typical non-circular coding, similar to one in
Henderson's book \cite{Hen80} is
\begin{verbatim}
 let knot p x = not(p x)
 and mult m n = n mod m = 0
 in let rec
     from n = n.(from (n+1))
 and sieve L =
      (hd L)
     .(sieve (filter (knot mult (hd L)) L))
 in sieve (from 2)
\end{verbatim}

This program creates many intermediate lists --
\texttt{from\ 2} and various sublists containing fewer and fewer composites.
This is due to the many successive calls on filter each
of which returns a list.

There are two main families of primes programs in imperative programming.
A sieve program finds successive primes and for each prime removes all
multiples of it from the set of numbers.
A program of the other family maintains a set of primes.
There is a loop over new candidates and each candidate
is tested for primality against members of the set.
It may then be added to the set.
The filtering of each candidate becomes the inner operation
and it can be coded as follows:
\begin{verbatim}
  let rec
     multiple L n =
        if sqr(hd L) > n then false
        else if mult (hd L) n then true
        else multiple (tl L) n
  in let rec
\end{verbatim}
\begin{verbatim}
     primes = 2
       . (filter (knot (multiple primes))
                 (from 3))
  in primes
\end{verbatim}
In this circular program, the expression for \texttt{primes}
is self-referential.
It starts with \texttt{2} and a sublist of \texttt{from\ 3} follows.
All composite numbers are removed by one call to filter
and so no intermediate lists are created.
The predicate \texttt{multiple\ primes} tests if a number is composite
by examining only \texttt{primes} already calculated.
When a new number is tested for primality, primes
exceeding its square root are known
and so it is not necessary to pass the number of known primes to multiple
(compare this with breadth-first and \texttt{unique}).

The central expression \texttt{knot\ (multiple primes)}
is precisely the predicate \texttt{isprime}.
This observation yields the alternative circular program:
\begin{verbatim}
  let rec
    primes = 2
           . (filter isprime (from 3))
  and isprime = knot (multiple primes)
  in primes
\end{verbatim}

\texttt{primes} and \texttt{isprime} form a mutually recursive
data structure and function pair.
Alternatively, substituting \texttt{primes} in \texttt{isprime} gives:
\begin{verbatim}
 let rec
     isprime = knot(multiple
                    (2.(filter isprime
                               (from 3))))
\end{verbatim}

Substituting and moving between these three programs
brings no disadvantage
except perhaps for the loss of the ability to refer to both
\texttt{primes} and \texttt{isprime},
because no circular data structure
is created in any of the programs.
The behaviour of \texttt{isprime} is interesting because it runs faster
the second time that it is called on a number of a given order
of magnitude because the primes that it needs have already been
calculated.

\section{Imperative Languages}

On circular programs, Bird \cite{Bir84} states
``the Pascal programmer confronted with the same idea
for optimization has to undertake a major revision of his or her
program to achieve the same end''.
It is certainly true that Bird's transformations, and many others, are
very hard to apply systematically in an imperative language because
of side-effects
but a circular program can often be written quite easily in
such a language.
If using an imperative language is a condition of the job,
a circular program might still be coded
to give an elegant and efficient result.

A direct translation of unique, for example, into Pascal
will \textit{not} work because Pascal uses strict evaluation.
But
\begin{verbatim}
  function f(...):...;
  begin ... f:=e; ... end
\end{verbatim}
can be
replaced by
\begin{verbatim}
  procedure f(... var fresult:...);
  begin ...; fresult:=e ... end
\end{verbatim}
which is defined more often than the former.
If this is applied to unique, the result is:
\begin{verbatim}
 function unique(L:list):list;
   var r :list;
   function member(e:element;
                   L:list;
                   n:integer):boolean; ...;

   procedure u(L:list; n:integer;
               var uresult:list);
   begin
     if L=nil then uresult:=nil
     else if member(L^.hd, r, n) then
       u(L^.tl, n, uresult)
     else
     begin
       uresult:=cons(L^.hd, nil);
                (* nil - not a recipe *)
       u(L^.tl, n+1, uresult^.tl)
     end
   end;
 begin r:=nil; u(L, 0, r); unique:=r
 end
\end{verbatim}

Note that the list \texttt{r} is \texttt{nil} terminated so a more usual
version of member can be adopted and parameter \texttt{n} of
routines \texttt{u} and \texttt{member} can be dropped.
Needless to say the Pascal version will not run on infinite lists.
A similar transformation can be performed to give a Pascal version
of the breadth-first traversal program and of the other programs.

The Pascal version of \texttt{unique} is simply doing what the implementation
of a lazy language would do for a circular program.
It marks the current end of the result \texttt{r} as
\texttt{nil} for unevaluated
and this is overwritten when its value is known.

One of Bird's examples is to transform a tree,
into a tree of the same shape but replacing
the leaf values by the minimum leaf value of the input tree.
His circular program builds a leaf with a minimum but
unevaluated value while incorporating this leaf in a tree
of the correct shape.
The leaf's value is calculated during the same traversal that copies the
tree's shape.
A Pascal version is almost as simple.
It creates a node, traverses the input tree incorporating the node
in a new tree and also searching for the minimum value.
When all is done this value is stored in the leaf.
Again this mimics the lazy implementation.

\section{Conclusion}

A circular program uses a recursive expression for a data structure.
In cases where the evaluation of the expression incorporates the data structure
directly, the result is a circular data structure.
Circular programs permit classic data structures such as
circular and doubly-linked lists, threaded trees, and queues to be
used in a functional programming language
and brings some of the efficiency of imperative programming
to functional programming.
Provided that the structures are subject to the single assignment rule,
reference variables and assignment (\texttt{:=}) are not needed.
Many times a circular program is more space-efficient than its conventional
counterpart by avoiding the creation of intermediate structures.

A lazy functional language is needed to write a circular program.
Garbage collectors based purely on reference counts
cannot collect circular structures
but mark-scan collectors,
copying collectors and hybrid schemes can \cite{Coh81}.
Since the programs are space efficient, the collector
should in any case be called infrequently.

In addition to the use of functional languages,
a circular program can often be translated into
an imperative language such
as Pascal with only minor revision of ideas.


\bibliographystyle{plainurl}

\bibliography{paper}

\noindent
\rule{\linewidth}{1pt}

\subsubsection*{This copy}

This is based on TR87/91 \cite{All87}
(as was
\begin{itemize}
\item[{[}10{]}] L. Allison.
  Some applications of continuations'
  \textit{Computer Journal}, 31(1):9-11,
  \myhref{https://doi.org/10.1093/comjnl/31.1.9}{doi:10.1093/comjnl/31.1.9},
  1988);
\end{itemize}
a later version appeared as
\begin{itemize}
\item[{[}11{]}] L. Allison.
  Circular programs and self-referential structures.
  \textit{Software Practice and Experience}, 19(2):99--109,
  \myhref{https://doi.org/10.1002/spe.4380190202}{doi:10.1002/spe.4380190202},
  1989.
\end{itemize}
More examples of circular programs can be found in
\begin{itemize}
\item[{[}12{]}] L. Allison.
  Applications of recursively defined data structures.
  \textit{Australian Computer Journal}, 25(1):14--20,
  \myhref{https://arxiv.org/abs/2206.12795}{arxiv:2206.12795},
  February 1993.
\end{itemize}
and at
\myhref{https://www.cantab.net/users/mmlist/ll/FP/Lambda/}{www.cantab.net/users/mmlist/ll/FP/}
\newline
\newline
-- L.A., orcid id
[\myhref{https://orcid.org/0000-0002-9020-3164}{0000-0002-9020-3164}].

\end{document}